\documentclass[twocolumn,twoside,slac_two]{revtex4}
\usepackage{graphicx}
\usepackage{fancyhdr}
\usepackage{amssymb}
\newcommand{\tHe}{\ensuremath{{}^3\!\mathrm{He}}}

\newcommand{\magHg}{\ensuremath{{}^{199}\mathrm{Hg}}}

\newcommand{\magXe}{\ensuremath{{}^{129}\mathrm{Xe}}}

\newcommand\geant{\textsc{Geant4}\/}
\newcommand\geantucn{\textsc{Geant4-UCN}\/}
\begin{document}

\title{An Improved Neutron Electric Dipole Moment Experiment}

%

\author{K. Bodek, St. Kistryn, \underline{M. Ku\'zniak}, J. Zejma}
\affiliation{JUC, Jagellonian University, Cracow, Poland}
\author{M. Burghoff, S. Knappe-Gr{\"u}neberg, T. Sander-Thoemmes, A. Schnabel, L. Trahms}
\affiliation{PTB, Physikalisch Technische Bundesanstalt, Berlin, Germany}
\author{G. Ban, T. Lefort, O. Naviliat-Cuncic}
\affiliation{LPC-Caen, ENSICAEN, Universit\'e de Caen, CNRS/IN2P3-ENSI, Caen, France}
\author{N. Khomutov}
\affiliation{JINR, Joint Institute for Nuclear Research, Dubna, Russia}
\author{P. Knowles, A.S. Pazgalev, A. Weis}
\affiliation{FRAP, Universit\'e de Fribourg, Fribourg, Switzerland}
\author{G. Rogel}
\affiliation{ILL, Institut Laue-Langevin, Grenoble, France}
\author{G. Qu\'em\'ener, D. Rebreyend, S. Roccia, M. Tur}
\affiliation{LPSC, Laboratoire de Physique Subatomique et de Cosmologie, Grenoble, France}
\author{G. Bison}
\affiliation{BMZ, Biomagnetisches Zentrum, Jena, Germany}
\author{N. Severijns}
\affiliation{KUL, Katholieke Universiteit, Leuven, Belgium}
\author{K. Eberhardt, G. Hampel, W. Heil, J.V. Kratz, T. Lauer, C. Plonka-Spehr, Yu. Sobolev, N. Wiehl}
\affiliation{GUM, Gutenberg Universit\"at, Mainz, Germany}
\author{I. Altarev, P. Fierlinger, E. Gutsmiedl, S. Paul, R. Stoepler}
\affiliation{TUM, Technische Universit\"at, M\"unchen, Germany}
\author{M. Daum, R. Henneck, K. Kirch, A. Knecht, B. Lauss, A. Mtchedlishvili, G. Petzold, G. Zsigmond}
\affiliation{PSI, Paul Scherrer Institut, Villigen, Switzerland}

\begin{abstract}
A new measurement of the neutron EDM, using Ramsey's method of separated oscillatory fields, is in preparation at the new high intensity source of ultra-cold neutrons (UCN) at the 
Paul Scherrer Institute, Villigen, Switzerland (PSI). The existence of a non-zero nEDM would violate both parity and time reversal symmetry and, given the CPT theorem, might lead 
to a discovery of new CP violating mechanisms. Already the current upper limit for the nEDM ($|d_n|<2.9\times10^{-26} e$\,cm) constrains some extensions of the Standard Model.

The new experiment aims at a two orders of magnitude reduction of the experimental uncertainty, to be achieved mainly by (1) the higher UCN flux provided by the new PSI source, 
(2) better magnetic field control with improved magnetometry and (3) a double chamber configuration with opposite electric field directions.

The first stage of the experiment will use an upgrade of the RAL/Sussex/ILL group's apparatus (which has produced the current best result) moved from Institut Laue-Langevin 
to PSI. The final accuracy will be achieved in a further step with a new spectrometer, presently in the design phase.
\end{abstract}

\maketitle

\thispagestyle{fancy}

\section{Motivation}
Although all experiments conducted so far measured a value consistent with 
zero, there are still reasons to suspect that the neutron has an electric dipole 
moment significantly larger than $10^{-31\pm1} e$\,cm~\cite{nedmSM}, predicted by the Standard Model (SM).

In general, the existence of
particles with an EDM would violate both parity ($\cal P$) and time reversal 
symmetries ($\cal T$). Invoking $\cal{CPT}$ invariance this is equivalent
to a violation of $\cal{CP}$ invariance, as discovered e.g. in the K$^0$ and B$^0$ systems.
Additional unknown sources of $\cal{CP}$ violation are of much interest as a 
possible explanation of one of the biggest puzzles in modern physics, namely, the
almost 9 orders of magnitude large discrepancy between the observed Baryon Asymmetry
of the Universe (BAU) and the one predicted by the SM and the Big Bang theory. 
$\cal{CP}$ violation is introduced with the $\delta$ phase in the 
Cabibbo-Kobayashi-Maskawa (CKM) mixing matrix and tuned such that it explains
the symmetry breaking observed in the K$^0$ and B$^0$ sectors.

As suggested by Sakharov \cite{sakharov}, the missing BAU could dynamically arise from an initial state 
with baryon number equal zero if the following conditions hold: (\romannumeral1) baryon number non-conservation, 
(\romannumeral2) the existence of both $\cal C$ and $\cal{CP}$ violating processes (\romannumeral3) occurring in a non-equilibrium state at an early 
epoch in the Universe. The discovery of the nEDM, as a clear indication of a new source of $\cal{CP}$ violation, could 
help to unravel the problem.

The most recent experimental result gives an upper limit of $|d_n| < 2.9
\times 10^{-26} e\,$cm~\cite{Bak06}. The nEDM search is at this point 
statistically limited and also facing systematic challenges not 
far away~\cite{Bak06, Pen04, Lam05, Bar06,Har06a}.  Further progress 
needs better statistical sensitivity and control of systematics.  
Several collaborations around the world aim at improved EDM experiments.

Interestingly, some theoretical extensions of the SM, including different types
of supersymmetric models predict $10^{-28}<|d_n|<10^{-25} e\,$cm~\cite{Abe01,Fal99}. 
Therefore, an improvement of the 
present upper limit by one or two orders of magnitude would be a
sensitive test of physics beyond the SM and could provide some
answers essential for our understanding of the Universe.

\section{Measuring Principle}
The measurement is made with neutrons stored in a cell (bottle) placed in uniform collinear $E$- and $B$-fields. 
The Hamiltonian determining the energy states of the neutron contains the terms 
$­{\bf \mu_n\,B}$ and ${\bf d_n\,E}$, where ${\bf \mu_n}$ denotes the neutron magnetic moment 
and ${\bf d_n}$ the hypothetical electric dipole moment.
Depending on the relative orientation (parallel or anti-parallel) of the $E$- and $B$-fields, the energy
of the state is given by  $h\nu_{\uparrow\uparrow} = 2|\mu_n|B - 2d_nE$ or 
$h\nu_{\uparrow\downarrow}= 2|\mu_n|B + 2d_nE$, respectively.
The precession frequencies $\nu$ relate to $d_n$ via 
\begin{equation}
h\delta\nu \equiv h(\nu_{\uparrow\uparrow} - \nu_{\uparrow\downarrow}) = -4d_nE.         
\end{equation}
Thus, the goal is to measure, with the highest possible sensitivity, the shift $\delta\nu$ when a strong 
$E$ field is reversed relative to the direction of $B_0$, the main magnetic field in the experiment.

Neutron storage experiments of this type are possible with ultracold neutrons (UCN), which have kinetic energies
of the order of 100\,neV and can be reflected under any angle of incidence from certain materials, like e.g. 
diamond-like carbon (DLC). The reflection is caused by the coherent strong interaction of the neutron with atomic 
nuclei. It can be quantum-mechanically described by an effective material-specific potential which is commonly 
referred to as the Fermi pseudo potential.

In the RAL/Sussex/ILL apparatus~\cite{Bak06} polarized UCN are introduced into the storage chamber and their 
precession frequency $\nu$ in the magnetic field $B_0$ and electric field $E$ is measured using the Ramsey separated 
oscillatory field method. The neutron spins are precessed by $\pi/2$ by a magnetic field pulse transverse to $B_0$ 
and oscillating at the neutron Larmor frequency, $B(t) = B_T\cos(2\pi\nu_L\cdot t)$. 
Then the  neutrons precess freely (around the direction of $B_0$) for a time $T$ ($\sim$130\,s) and, if the neutron electric 
dipole moment is non-zero, $d_n \neq 0$, the precession frequency in the combined magnetic and electric field is different
from the Larmor frequency. 
Thus, a phase difference proportional to the precession time $T$ builds up, $\phi\approx(2\pi\nu - 2\pi\nu_L)\cdot T$, 
which has opposite sign for $E\uparrow$ and $E\downarrow$. During time $T$, $B_0$ is monitored by a $^{199}$Hg co-magnetometer, 
i.e. polarized mercury vapor stored in the same volume with the UCN.
After that the oscillating field is activated again (strictly in phase with the first $\pi/2$ flip)
and the UCN are again precessed by $\pi/2$.
If $d_n = 0$, the spins after two $\pi/2$ pulses are all oriented anti-parallel to their initial direction.
In any other case, the accumulated phase shift results in a different neutron spin orientation. 
The last step is 
to analyze the number of neutrons $N_{up}$ and $N_{down}$ that finish in the two spin states (up or down) relative to $B_0$. 
This is accomplished by transmission through a magnetized iron foil, the same which is used for polarizing when filling the 
chamber.

The experiment is operated on a batch cycle principle: (a) fill with polarized neutrons, (b) carry out the magnetic resonance, 
and (c) empty, spin analyze and detect to obtain $N_{up}$ and $N_{down}$. 
The cycles are conducted 
continuously, while the direction of $E$ is reversed a few times per day. 
The error due to counting statistics is given as
\begin{equation}\label{eq-errdn}
\sigma(d_n) = \frac{\hbar}{2\alpha ET\sqrt{N}},         
\end{equation}
where $N$ is the neutron counts and $\alpha$, called visibility, represents the efficiency of maintaining 
the polarization throughout the measuring sequence.

Equation~\ref{eq-errdn} can be applied assuming that the $B_0$ field does not change over the series of measurements. This assumption is to a large extent 
fulfilled, because of the magnetic shield, which suppresses the ambient field. Residual changes of the magnetic field can still be corrected 
for with the $^{199}$Hg magnetometer. 

\section{Precision Goals and Strategy}
The nEDM collaboration\footnote{http://nedm.web.psi.ch} was given access to the 
old RAL/Sussex/ILL~\cite{cryoedm} group's apparatus located at ILL, the one which was used to measure 
the currently best limit. The collaboration plans to further advance the (proven) in-vacuum room temperature technique of nEDM measurements, 
using the existing spectrometer in the initial phase of the project. The sensitivity improvement will be achieved mainly due to the high UCN flux at the 
new source at PSI\footnote{http://ucn.web.psi.ch} 
(two orders of magnitude improvement over the present ILL source). Main developments leading to better control over systematic 
effects contain additional magnetometry systems and better magnetic shielding and stabilization.

In more detail, the PSI project consists of three phases:
\begin{itemize}
\item Phase I (at ILL, in progress during 2008): 
Improving the old apparatus, R\&D. 
\item Phase II (at PSI, 2009 -- 2011): 
Moving the apparatus to PSI and doing a measurement with $5\times10^{-27}e\,$cm sensitivity level. In parallel, 
design and construction of a new double-chamber nEDM spectrometer. 
\item Phase III (at PSI, 2011 -- 2015): 
  Measurement with the new apparatus aiming at $5\times10^{-28}e\,$cm sensitivity. 
\end{itemize}

\section{Phase I: Ongoing R\&D}
While the PSI UCN source is under construction we have concentrated
on tests and improvements of the RAL/Sussex/ILL apparatus at ILL Grenoble.
\subsection{Simulations}
\begin{figure}[h]
\centering
\includegraphics[width=85mm, height=70mm]{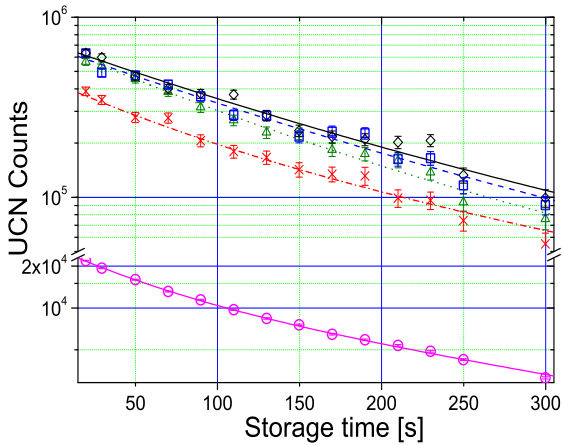}
\caption{Simulated neutron storage curves in the existing nEDM apparatus, 
  placed 1~m above the
  PSI UCN beamline: quartz insulator + DLC-coated electrodes (red $\times$, dash-dotted line), 
  DPS-coated insulator + DLC-coated electrodes (green $\triangle$, dotted line), 
  diamond coated insulator + DLC electrodes (blue $\Box$, dashed line), 
  diamond coated insulator and electrodes (black $\diamondsuit$, solid line). 
  Experimental curve from ILL shown for comparison (magenta $\circ$, 
  solid line)~\cite{thesis}. 
	} \label{fig:oillstor}
\end{figure}

We have simulated the situation in which the spectrometer is mounted at
the PSI source (with a superconducting polarizer in the horizontal
beamline), using a realistic simulation of the source and the
apparatus. It is performed using a special version of the CERN program 
\geant{} into which UCN particles and their
interactions have been implemented~\cite{geant4ucn}.
The results are displayed in Fig.~\ref{fig:oillstor} and show that 
the number of UCN stored (after about 150~s) can be increased
considerably if the Fermi potential of the chamber walls is changed
from 90~neV (quartz) to 162~neV (deuterated polystyrene,
DPS) or 304~neV (diamond).

In other simulations, \geantucn{} is being used to study possible
simultaneous polarization analysis~\cite{gwendal} and a velocity sensitive UCN detector: 
the motivation for the latter being the fact that various systematic 
false effects can have a dependence on
UCN velocity and call for proving their absence in a possible signal.
A dedicated external simulation, using the UCN trajectories from \geantucn\
as input, was created to study this sort of effects~\cite{knecht}.

Another necessary input are electric and magnetic field maps, created
with available commercial finite element 
codes or a custom-made tool for finite volume integrals. 
The field models are also
important for the design of a new 5-layer magnetic shield for 
Phase III of the project.

\subsection{New materials}
Simulations for the present setup attached to the new PSI source indicate
a factor of $\sim$20 gain in UCN counts after 150~s storage and another 
factor of 1.5,
when a material with a Fermi potential similar to DPS 
is used as a coating on the insulator walls. The gain obviously depends on the 
UCN energy spectrum. The Fermi potential of DPS was measured by cold neutron
reflectometry to be $162\pm2$~neV, in agreement with
expectation and a former determination by Lamoreaux~\cite{Lam88}. 

\begin{figure}[h]
\centering
\includegraphics[width=85mm, height=70mm]{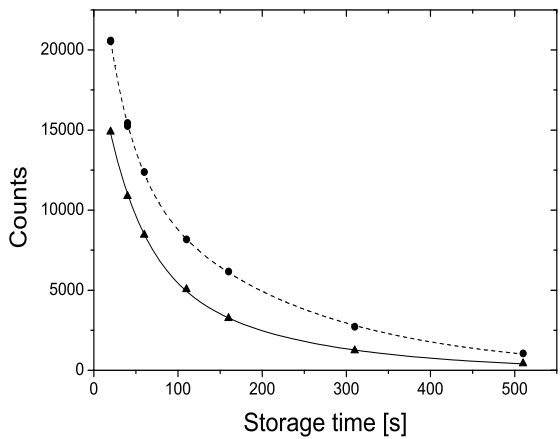}
\caption{Total number of neutrons vs. storage times measured for the quartz insulator ($\triangle$) 
    and the DPS-coated insulator ($\bullet$). DPS measurements were made two days after the 
    quartz results had been obtained. 
	}
\label{fig:dps}
\end{figure}

We have made a first test of a full-scale PS insulator ring ($\phi\sim50\,$cm) coated with DPS at the ILL 
source and obtained roughly 50\% increase in UCN counts after 150~s storage time (Fig.~\ref{fig:dps}). The result
is in agreement with simulations for the ILL source and additionally validates the gain factors 
predicted in a similar way for the new PSI source (Fig.~\ref{fig:oillstor}).

For the test of the PS ring we successfully employed UV-grade quartz windows coated by DPE (using
spin-coating to obtain an optically flat surface), since DPS
deteriorates under UV illumination, necessary for the co-magnetometer.   
The Fermi potential of DPE was also measured by means of cold neutron
reflectometry, yielding $V_F=212\pm2$~neV. 

The performance of the new chamber in terms of the neutron storage and depolarization,
high-voltage stability, electrical resistivity and compatibility with the  
$^{199}$Hg magnetometer is similar or better than for a quartz chamber~\cite{thesis}.
\subsection{Magnetometry}

In order to investigate the possibility to improve the \magHg{}
co-magnetometer sensitivity a detailed theoretical model, based on 
actual geometry of the system, has 
been developed, which aims at 
calculating the
parameters,
which affect the 
precision of the measurement.  
Based on the model, 
possible improvements have been identified and are being pursued~\cite{stephanie}.

In order to better control systematic effects, the magnetic field
and its gradients will be monitored and stabilized using an array of
laser optically pumped Cs-magnetometers~\cite{Weis05g}.
For the external magnetometry, a system of Cs magnetometers around the
UCN chambers (and ideally delivering vector information on the
magnetic field, as in principle possible from newly developed
double-resonance alignment magnetometers~\cite{DRAMexp}) will be used for three tasks: 
i)
stabilization of the magnetic field, ii) stabilization of the magnetic
field gradient(s), \romannumeral3) monitoring of the magnetic field.  
The first basic version of the system is already operational.

\subsection{Detection system}
The higher UCN density which will be available at the PSI source calls 
for new fast detectors able to cope with higher rates.
Recently the activities have progressed along two main
directions: 1) the comparative study of UCN detector performances,
with emphasis on their efficiency as well as on their background
sensitivities; and 2) the test of geometries for the simultaneous
analysis and detection of the two UCN spin components \cite{gwendal}.

An extensive comparison between Cascade-U 
(GEM-type) detectors and $^6$Li doped glass scintillators (GS3, GS10, GS20) indicated
higher efficiency of the scintillators for velocities in the range
from 6 to 12 m/s.

\begin{figure}[h]
\centering
\includegraphics[width=50mm]{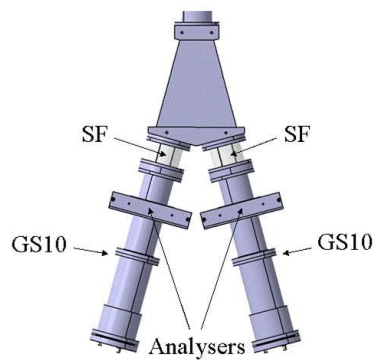}
\caption{Spin analysis system with a 30 degrees angle between the two arms (see text).} \label{fig-y30}
\end{figure}
The prototype polarization analysis system consists of two arms,
each of them equipped with an adiabatic spin-flipper (SF), a spin analyzer 
(magnetized iron foil) and a detector (Fig.~\ref{fig-y30}).
The transmissions integrated over the whole UCN velocity distribution 
have been found to be around $46\%$ for both arms and are velocity 
independent in the range from 4 to 9~m/s.  
The asymmetry between both arms, obtained from the counting rates with 
the respective spin flipper ON and OFF, measured for the UCN velocities 
ranging from 4 to 7~m/s, is close to 80\% (with the actual beam polarization
around 90\%). 
Further optimization of the system is planned.

\section{Phase II: Measurement at PSI}
\begin{figure}[h]
\centering
\includegraphics[width=80mm]{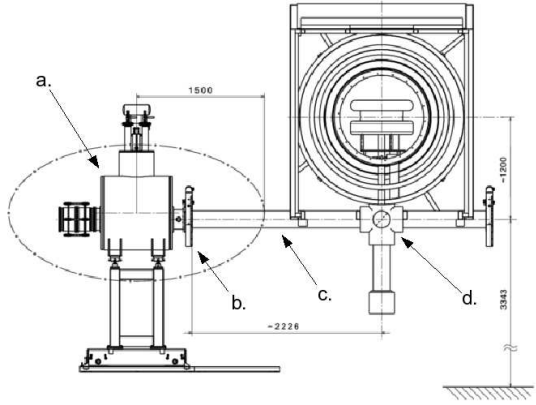}
\caption{A schematic of the former RAL/Sussex/ILL apparatus moved to PSI with 
    SC polarizer magnet (a), VAT valve (b), UCN guide system (c), UCN switch (d): spectrometer
    (up), detection system (down) and test beam port (straight).  The
    ellipse around the magnet indicates the 1~Gauss line.} \label{fig-oill}
\end{figure}

The concept for the Phase II setup is shown in
Fig.~\ref{fig-oill}.  From left to right, the UCN pass
through the superconducting polarizer magnet, a custom
designed VAT UCN vacuum valve, a UCN switch valve with its connections 
up into the UCN trap, down into the detection system and horizontally 
through-going to a second VAT UCN valve. The maximum number of UCN is 
obtained by minimizing the vertical distance between the storage chamber 
and the input guide.  
Not shown here are the monitor detector system in the fill
line, the vacuum system, the thermal housing and a possibly required
compensation coil system.

A sensitivity increase by a factor of $~5$ is anticipated due to the higher UCN intensity 
and better systematic control (based on the results of present R\&D, see the previous section). 

\section{Phase III: Double Chamber Setup}
The next generation setup will have a vertical field configuration 
and cylindrical, vertically stacked double UCN chambers inside a horizontal,
cylindrical, multilayer $\mu$-metal shield. Additional gain in sensitivity
is expected due to a larger trap volume ($\times\sqrt{3}$), better
adaptation to the UCN source  ($\times\sqrt{3}$), longer running time
($\times\sqrt{3}$) and electric field increase ($\times2$).

Better control over systematics will be secured by the advantages of the
double setup itself, possible velocity sensitive UCN detection 
and by the improved control over magnetic field, provided by the new shield, 
a field stabilization system and enhanced magnetometry, described below in more detail.  
\begin{figure}[h]
\centering
\includegraphics[width=80mm]{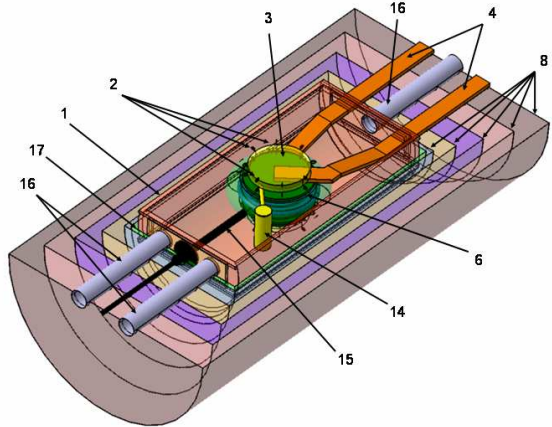}
\caption{A view of our presently ongoing design
  study for n2EDM\@.  The numbers label vacuum chamber (1), Cs
  magnetometer array (2) for both, B-field measurement and \tHe{}
  read-out, the two large \tHe{} magnetometer vessels (3), rectangular
  shaped UCN guides (4), UCN double-chamber (6), 5-layer $\mu$-metal shield
  (8), a co-magnetometry system (14), HV connection (15), various
  feedthroughs for pumping and electrical, optical and gas connections
  (16), and the inner coil system (17).} \label{fig-n2edm}
\end{figure}

\subsection{Magnetometry}
Magnetometry-wise we plan to combine co-magnetometry (improved
\magHg{} magnetometry, R\&D on \magXe{} and \tHe), with field control
by multiple Cs-sensors and two large volume \tHe{} magnetometers, also
read by Cs magnetometers.

\begin{figure}[h]
\centering
\includegraphics[width=80mm]{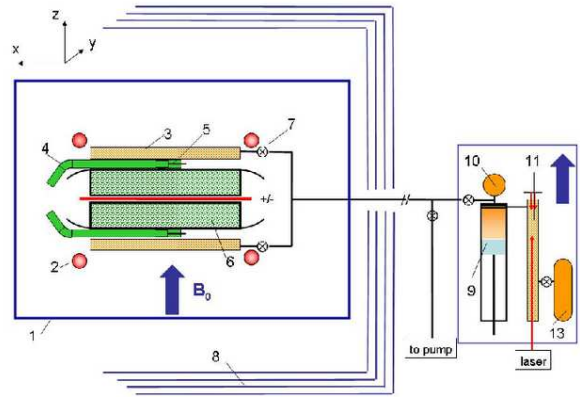}
\caption{Conceptual lay-out of the \tHe{} system for n2EDM~\cite{heil}. See text
for details.} \label{fig-n2he}
\end{figure}

The \tHe{} magnetometry setup is displayed in
Fig.~\ref{fig-n2he}, with vacuum chamber (1), Cs magnetometer
array (2) for both, B-field measurement and \tHe{} read-out, 2 large
\tHe{} magnetometer vessels (3), rectangular shaped UCN guides (4),
valves (5) and double-chamber (6).  The Cs magnetometer array will
have multiple sensor heads (of order 64--128, not shown) and a
co-magnetometry system (not shown) based on at least one out of three
R\&D options (\magHg, \magXe, \tHe) will be used.

Initially, \tHe{} is provided by a reservoir (13).  Spin-polarized
\tHe{} is produced by the method of metastable optical
pumping~\cite{Eck92} in a cylindrical optical pumping cell (11) where
a weak gas discharge is maintained to populate the metastable
(${}^3$S$_1$) states of \tHe{}.  Following polarization, the gas at
around 1~mbar is compressed into a buffer-cell (10) by means of a
nonmagnetic piston compressor (9).  This polarization and compression
cycle is repeated until a pressure of about 100~mbar is reached in the
buffer-cell. Then the gas is expanded via transfer-lines and suitable
valves (7) into the magnetometer vessels.  It should be noted that the
whole \tHe{} pumping and compression unit is outside the multilayer
$\mu$-metal shield (8), in order not to disturb the electric and
magnetic field conditions.

Our feasibility studies show that the proposed \tHe{}
magnetometer is capable of tracing tiny magnetic field variations of
$\delta B\approx2$~fT during a typical UCN storage cycle of 200~s
and can additionally monitor the gradients with high precision~\cite{heil}.
\section{Summary}
We propose improving
the in-vacuum technique and performing a competitive experiment in
steps, delivering first results in about 4 years (i.e., 2 years after
set-up at the running PSI UCN source) with a five-fold improved
sensitivity to a level of $5\times 10^{-27} e\,$cm and reaching a
sensitivity of $5\times 10^{-28} e\,$cm in 6--8 years.

Several collaborations around the world aim
at improved EDM experiments.
The new cryogenic EDM searches~\cite{cryoedm, Ito07, snsedm, Pen06}
aim at producing high UCN densities in super-fluid helium and
performing also the EDM experiment directly in this medium. 

In view of the experimental challenges of more precise EDM
experiments, the competition of different groups using varying,
complementary techniques could be crucial.  While being on a similar
time scale and with comparable sensitivity, our
approach uses a different technique and offers unique features for
systematic control, especially with respect to magnetometry.

\end{document}